\documentclass[preprint,showpacs,showkeys,amsmath,amssymb]{revtex4}

\usepackage{graphicx}% Include figure files
\usepackage{dcolumn}% Align table columns on decimal point
\usepackage{bm}% bold math

\begin{document}

\preprint{}

\title{Non-Parametric Extraction of Implied Asset Price Distributions}

\author{Jerome V Healy}
\author{Maurice Dixon}
\altaffiliation[]{Visitor: e-Science, RAL, Didcot, Oxon, OX11 0QX}
\email{M.Dixon@Londonmet.ac.uk}
\author{Brian J Read}
\author{Fang Fang Cai}
\affiliation{  CCTM, London Metropolitan University, 31 Jewry Street, London EC3N 2EY
}

\date{\today}% It is always \today, today,

\begin{abstract}

Extracting the risk neutral density (RND) function from option prices is well defined in principle, but is very sensitive to errors in practice.  For risk management, knowledge of the entire RND provides more information for Value-at-Risk (VaR) calculations than implied volatility alone [1]. Typically, RNDs are deduced from option prices by making a distributional assumption, or relying on implied volatility [2].  We present a fully non-parametric method for extracting RNDs from observed option prices.  The aim is to obtain a continuous, smooth, monotonic, and convex pricing function that is twice differentiable.  Thus, irregularities such as negative probabilities that afflict many existing RND estimation techniques are reduced.  Our method employs neural networks to obtain a smoothed pricing function, and a central finite difference approximation to the second derivative to extract the required gradients.

This novel technique was successfully applied to a large set of FTSE 100 daily European exercise (ESX) put options data and as an Ansatz to the corresponding set of American exercise (SEI) put options.  The results of paired t-tests showed significant differences between RNDs extracted from ESX and SEI option data, reflecting the distorting impact of early exercise possibility for the latter.  In particular, the results for skewness and kurtosis suggested different shapes for the RNDs implied by the two types of put options.  However, both ESX and SEI data gave an unbiased estimate of the realised FTSE 100 closing prices on the options' expiration date.  We confirmed that estimates of volatility from the RNDs of both types of option were biased estimates of the realised volatility at expiration, but less so than the LIFFE tabulated at-the-money implied volatility. 
\end{abstract}

%Use showkeys class option if keyword
\maketitle

\section{\label{sec:level1} Introduction}

Many asset pricing models used in finance, including the Black-Scholes (BS) model for option prices, rely on the conventional assumption that the statistical distribution of asset returns is normal, and the price distribution log-normal.  This assumption is consistent with geometric Brownian motion as the underlying mechanism driving price movements.  It is now well known that historical asset price distributions exhibit fat tails. That is, they are slightly smaller near the mean and larger at extreme values.  This has important implications for financial risk management, as it suggests that large price movements occur more frequently than they would for a normal distribution with the same variance.  It also suggests that the underlying price process does not follow a geometric Brownian motion. Option prices represent a rich source of information on the statistical properties of the underlying asset. Exchange traded options are now available on financial assets including stock indices and futures; often these are heavily traded, so are very liquid. While daily time series of asset prices contain just one observation per day, there is a set of option prices available for each maturity date. These option prices reflect traders' expectations regarding future price movements of the underlying asset so they allow alternative approaches to estimating financial risk. 

In this paper we present a simple yet effective method for extracting non-parametric estimates of the complete distribution for the value of the underlying asset at maturity of an option - known as the  risk neutral density (RND)- from sets of daily option prices. We have recently reported the application of data mining techniques using neural nets to model European style FTSE 100 index options [3]. We extended that approach by introducing prediction intervals [4]. Here, we demonstrate how neural nets can be used to obtain estimates of the expected value and standard deviation (volatility), as well as higher moments, of the implied distribution of the asset price at the expiration of an option written on it.

%The paper is organised as follows. In Section II the theory, and existing methods for extracting RNDs are introduced. The modelling approach and the data used are presented in Section III. Results are reported in Section IV for the application of the approach to ESX, European put options. We then consider the consequences of applying it to SEI, American put options. Section V contains a discussion and conclusions.

\section{ Risk Neutral Distributions}
RNDs have many practical applications. They are used by central banks to assess market expectations regarding future stock prices, commodity prices, interest rates, and exchange rates in connection with setting monetary policy [2]. They are useful to market practitioners as an aid to investment decisions. RNDs extracted from exchange traded options can be used to price exotic options.  For risk management, they can provide measures of value-at-risk [1]. 

The prices of European exercise options can be expressed as the expected value of their payoffs, discounted at the risk-free interest rate.
\begin{subequations}
\label{eq:whole}
\begin{equation}
 C(X,t,T) = e^{ - r(T - t)} \int\limits_X^\infty  {\rho (S_T )(S_T  - X)dS_T} ,\label{subeq:1}
\end{equation}
\begin{eqnarray}
 P(X,t,T) = e^{ - r(T - t)} \int\limits_{ - \infty }^X {\rho (S_T )(X - S_T )dS_T } .\label{subeq:2}
\end{eqnarray}
\end{subequations}
In equation (1) {\textit {C(X,t,T)}} and {\textit {P(X,t,T)}} are the prices of calls and puts trading at time {\textit {t}} for expiration at some later time {\textit {T}}.  {\textit {X}} is the strike price, and {\textit {r}} is the risk-free interest rate. $\rho (S_T )$ is the distribution for the value of the underlying asset {\textit {S}} at {\textit {T}} predicted from time {\textit {t}}.  Given an assumption about the functional form of  $\rho (S_T )$, options can be priced for any value of exercise price {\textit {X}}.  Conversely, given a series of synchronous market prices observed at some time {\textit {t}}, for options expiring at some later time {\textit {T}}, this calculation can be inverted and an estimate of $\rho (S_T )$  extracted.  Breeden and Litzenberger [5] showed that the cumulative density function (negatively signed) for the value of the underlying asset {\textit {S}} at time {\textit {t}} is given by the first partial differential with respect to {\textit {X}} of equation (1), while the RND is obtained by differentiating equation (1) twice with respect to {\textit {X}}.
\begin{equation}
%{\textstyle{{\partial f(X,t,T)} \over {\partial X}}} =  - e^{ - r(T - t)} \int\limits_X^\infty  {\rho (S_T )dS_T %},and \hspace{6pt}  %
{\textstyle{\frac{\partial f(X,t,T)}{\partial X}}} =  - e^{ - r(T - t)} \int\limits_X^\infty  {\rho (S_T )dS_T }, and \hspace{6pt}  %
%\end{equation}
%\begin{equation}
{\textstyle{\frac{ \partial ^2 f(X,t,T)}{\partial X^2 }}} = e^{ - r(T - t)} \rho (S_T ). %
\end{equation}
In equation (2),  ${f(X,t,T)}$ represents the call or put European option pricing functions. In reality, {\textit X} is not continuous and options are only available for a limited number of exercise prices at discrete intervals. It has been shown [5], that for discrete data, finite difference methods can be used to obtain a numerical solution to equation (2). Neuhaus [6] has shown how the RND can be obtained via equation (2) using finite differences on the first derivative.

\subsection{\label{sec:level2}Recovering RNDs: Existing Methods}

Two techniques have been adopted as standard by practitioners.  They are the mixture of lognormals method and the smoothed implied volatility smile.  The first of these is a parametric method which works with equation (1); the second is a non-parametric method which works with equation (2).

The mixture of lognormals technique originated with Ritchey [7].  In this method, the RND is represented by the weighted sum of independent lognormal density functions.
\begin{equation}
\rho (S_T ) = \sum\limits_{i = 1}^m {[\Theta _i LnD(\alpha _i ,\beta _i ;S_T )],  }  \hspace{6pt} \sum\limits_{i = 1}^m {\Theta _i  = 1, }  \hspace{6pt} \Theta _i  > 0\forall _i . %
\end{equation} 
In equation (3), $ { LnD(\alpha _i ,\beta _i ;S_T )}$ is the i${^{th}}$ lognormal density function for the asset price at maturity $S_T$ , in the m component mixture with parameters $\alpha _i$ and $\beta _i$.  The parameter $\Theta _i$ is a probability weighting. This representation has the advantage that it offers greater flexibility than a single density representation.  In principle, a mixture of Gaussian (or other) densities can approximate any continuous density to arbitrary accuracy as the number of component density functions tends to infinity.  Ritchey used a mixture of two lognormal densities to minimise the number of parameters to be estimated, and this has become the standard procedure. If equation (3) is substituted into equation (1) the resulting expressions can be fitted to observed call and put prices, and the parameters estimated to minimise the weighted sum of fitted price errors, using non-linear optimisation methods.  Given the parameters and the observed option prices, the implied RND can then be constructed. In  this method the number of parameters to be estimated increases rapidly with the number of density functions included. The method is prone to overfitting since option price series frequently have 20 or fewer observed prices corresponding to different exercise prices with the same {\textit t} and {\textit T}.
 
The smoothed implied volatility smile method originated with Shimko [8].  The RND can be obtained directly from equation (2) provided the option pricing function $f(X,t,T)$ is observable.  Unfortunately, only a relatively small number of option prices corresponding to discrete exercise prices are observable for a given time {\textit t}.  An obvious solution is to smooth and interpolate the observed prices by fitting a function to them.  Shimko considered this, but found that his smoothing spline functions were not suitable for fitting option prices.  To overcome this difficulty Shimko converted the prices to implied volatilities using the BS formula.  He then fitted a quadratic polynomial smoothing function to the available implied volatilities and used linear extrapolation outside the range of observable exercise prices.  The continuous implied volatilities obtained were then converted back to continuous option prices using BS, and an RND extracted using the relation in equation (2).  In this method, the BS formula is used purely to effect a transformation from one data space to another; it does not rely on any of the assumptions underlying the BS formula. The smoothed volatility smile method has the advantage that fitting polynomial curves or splines to the smile can be done in a single pass, without iteration.  However, the probabilities in the resulting RND cannot be guaranteed to be positive because this is not a constrained optimisation.

Bliss and Panigirtzoglou [9] found that the smoothed implied volatility smile estimation method produced more stable RNDs than the mixture of lognormals method. By contrast, the mixture of lognormals method was found to be sensitive to computational problems, suffered from frequent convergence failures, and had a tendency to produce spurious spikes in estimated RNDs, hence they suggest it should not be used. Additionally, they suggest the smoothed implied volatility smile method may be improved by trading goodness-of-fit for a more stable RND, and remark that the mixture of lognormals method does not permit this fine tuning.  They warn though, that skewness and kurtosis cannot always be precisely estimated with either method. However, the disadvantages of the smoothed implied method include: 1) the need to convert prices to and from implied volatility, 2)the use of a limited family of quadratic functions to fit the data, 3) difficulty in obtaining a smooth join when extrapolating outside the range of observable exercise prices. 

\subsection{Recovering RNDs from Index Options with Neural Nets }
Surprisingly, there has been little study of the use of neural nets (NNs) for extracting RNDs from option prices.  A version of the parametric mixture of lognormals method was implemented by Schittenkopf and Dorffner [10] using Mixture Density Networks for the FTSE 100. Herrmann and Narr [11] differentiated NN option pricing functions fitted directly to prices of options on the German DAX index to obtain RNDs.  They used average values for some input variables when training their models.  The resulting RNDs were compared graphically with corresponding lognormal RNDs obtained using the BS formulae.  No statistical tests were performed, and only goodness-of-fit and error measures were provided.  Despite an extensive literature search, no other studies were found. The non-parametric extraction of RNDs from option prices is an example of an ill-posed problem, in that small changes in option prices can lead to large changes in the estimated density function.  It requires the use of methods that are robust to slight changes in the option prices. Neural nets have been shown to be suitable for directly fitting option prices, avoiding the need to work in implied volatility space as in the smoothed implied volatility method.

\subsection{American Put Options}
Most exchange traded options are American options and these can be exercised at any time prior to maturity. It is expected that this feature, which is reflected in their price, affects any extracted RND. The theory underlying RNDs is only applicable to European exercise options, and cannot be applied to American exercise options without modification. Dupont [12] discussed this and suggested that the early exercise correction is not significant in practice except for deep in-the-money options. It is an open question whether RNDs extracted from American options are significantly different empirically from those extracted from corresponding European options. The American style SEI option is based directly upon the FTSE 100 index while the European ESX option is based upon an implied future on the FTSE 100 index. The  underlyings converge when time approaches maturity date. Only put options are considered here; in the absence of dividend payments early exercise of American call options is never optimal, so they can be priced as European. Here, RNDs from American Ansatz and European put option pricing functions are extracted and compared.

\section{ Data And Method}

For this work, smoothed prices corresponding to each exercise price in a daily price series were estimated through directly fitting a neural net to create an option pricing function. Then RNDs were extracted by twice partially differentiating the functions numerically with respect to exercise price. A training set of 13,790 FTSE 100 ESX put options was prepared along with another training set which comprised 14,619 FTSE 100 SEI put options. To ensure prices were liquid, only options with positive values for contract volume and open interest were selected.  A disjoint test set of 60 daily option price series, containing data on a total of 1,238 (European) put options on the FTSE 100 index, was created. Option and underlying asset prices for the American puts were added for the same time, risk-free interest rate, and interleaved exercise price. The resulting test set had the following two special features: 1) Included options were traded for one of 60 consecutive monthly expirations, 2) The options had a maturity of one calendar month (17 or 18 trading days).  This was because any longer maturity results in overlapping data for some variables.

In creating the test set, the objective was to obtain a set of option price series with constant maturities, which was non-overlapping.  The latter feature was required to avoid serial dependence between successive observations, which might bias statistical results for the RNDs [13].  Pricing models for European and American put options were separately trained, using logistic functions in a 5-11-1 architecture.  The inputs were the five BS variables, and the targets were market prices of European and American put options, respectively. Once trained, the pricing models were applied to the test set to generate series of smoothed European and American option prices, taking care to use the correct values of the underlying asset as inputs to each model.  Each generated price series was then numerically differentiated to estimate ${\textstyle{\frac{\partial ^2 f(X,t,T)}{\partial X^2 }}}$  using symmetric central finite differences.  Making ${\textit f}$ correspond to the neural net, the following formula was applied:
\begin{equation}
\frac{{\partial ^2 f(X,S,t,r,\sigma )}}{{\partial X^2 }} = \frac{{f(X + \varepsilon ,S,t,r,\sigma ) - 2f(X,S,t,r,\sigma ) + f(X - \varepsilon ,S,t,r,\sigma )}}{{\varepsilon ^2 }}. %
\end{equation}
where $f(X,S,t,r,\sigma )$ is a neural net option pricing function with the 5 standard BS input variables defined in [3] using at-the-money volatilty, and $\varepsilon$ is a small increment.  If  $\hat \rho (X_i )$ is the RND  for a strike price interval $ [X_i ,X_{i+1}]$, then
\begin{equation}
\hat \rho (X_i ) \approx \frac{{\partial ^2 f(X,S,t,r,\sigma )}}{{\partial X^2 }}\varepsilon e^{ - r(T - t)}.% 
\end{equation} 
where $\varepsilon$ is the interval between adjacent values of ${X_i}$. Equation (5) was used to obtain point estimates of the RND for the interval corresponding to each $X_i$. The median value, and probabilities in the tails of the distribution below the $1^{st}$ and above the $n^{th}$ observable option price in the series, were estimated as suggested by [6], using equation (2).   
\begin{equation}
Mean = \sum\limits_{i = 1}^n {\frac{{X_i  + X_{i + 1} }}{2}} \hat \rho (X_i ).%
\end{equation}   
\begin{equation}
Stdev = \sqrt {(\sum\limits_{i = 1}^n {\frac{{X_i  + X_{i + 1} }}{2} - } Mean)^2 \hat \rho (X_i )}.% 
\end{equation}     
\begin{equation}
Skewness = (\sum\limits_{i = 1}^n {\frac{{X_i  + X_{i + 1} }}{2} - } Mean)^3 \frac{{\hat \rho (X_i )}}{{Stdev^3 }}.%
\end{equation}  
\begin{equation}
Kurtosis = (\sum\limits_{i = 1}^n {\frac{{X_i  + X_{i + 1} }}{2} - } Mean)^4 \frac{{\hat \rho (X_i )}}{{Stdev^4 }}.%
\end{equation} 
The first four moments of each recovered RND were estimated from Equations (6) to (9). The annualised percentage implied volatility was derived from equation (7) as 100*(Stdev/Mean)*$\nu^{0.5}$ where $\nu =$ (time-for-year/time-to-expiry). 

\section{Results: RNDs Of European And American Put Options}
The fit of the NN models to the market prices of the options in the test set gave an $R^2$ of  0.988 and 0.984 for the European and American put options respectively.  The calculated t-statistics were -0.09 and -1.32 respectively, compared with a critical value of 1.96 for a two tailed test.  These results confirmed that both NN models were unbiased estimators of market prices. Table I and Table II compare summary statistics for RNDs recovered from the test data of non-overlapping, constant maturity (1 month as 17/18 trading days), sets of European and American exercise put options on the FTSE 100 Index.
Table I gives results for direct comparisons of the time series of moments. The results of the paired t-tests indicate significant differences, for each statistic tested, between RNDs for European and American put options in spite of $R^2$=0.9996 for mean and median. In particular, the results for skewness and kurtosis suggested the shapes of the RNDs were different for each type of put option.  The F test is a test to determine whether two samples are from populations with different variances; here it indicates significant differences for annualised IV and for kurtosis.
\begin{table}
\scriptsize
\caption{\label{tab:table1}European v American FTSE 100 Index Put Options: Comparison of RNDs.}
\begin{ruledtabular}
\begin{tabular}{ccccc}
 $Summary Stat$ &$R^2$ &$F-stat$\footnotemark[1]
 &$t-stat(paired)$\footnotemark[1]  &$H0:$\footnotemark[1]\\
\hline
Mean& 0.9996 & 1.01 & 9.09 &Reject\\
Median &0.9996 & 1.00 & 5.07 &Reject\\
Stdev& 0.640 & 1.45 & 5.35 &Reject\\
Ann.I.V.$\%$ & 0.820 & 1.78 & 4.29 &Reject\\
Skewness& 0.932 & 1.09 & 19.01 &Reject\\
Kurtosis& 0.779 & 1.96 & -13.04 &Reject\\
\end{tabular}
\end{ruledtabular}
\footnotetext[1]{$F_{crit} =1.54.\hspace{3 pt} t_{crit}(2 \hspace{3 pt} tail) =2.00.\hspace{3 pt} H0:=no \hspace{3 pt} difference \hspace{3 pt} in \hspace{3 pt} SummaryStat \hspace{3 pt} means \hspace{3 pt} at \hspace{3 pt} 95\% \hspace{3 pt} confidence. $} %\footnotetext[2]{$t_{crit}(2 \hspace{6 pt} tail) =2.00.$}
%\footnotetext[3]{$H0:=no \hspace{6 pt} difference$}
\normalsize
\end{table}
\begin{table}
\scriptsize
\caption{\label{tab:table2}Performance: Comparison of Market Value with RNDs' One Month Estimate.}
\begin{ruledtabular}
\begin{tabular}{cccccc}
 $Option$ &$Parameter$ &$R^2$ &$F-stat$\footnotemark[1]
 &$t-stat$\footnotemark[1]  &$H0:$\footnotemark[1]\\
\hline
% &\multicolumn{3}{c}{$RNDs from ESX-European Exercise$}&\multicolumn{1}{c}{$$}\\
%RNDs from ESX 
ESX &T.FTSE100 v Median &0.955 & 1.01 & -0.02 &Accept\\
ESX &Realised Vol. v Ann.I.V.$\%$ & 0.379 & 2.01 & -3.47 &Reject\\
% &\multicolumn{3}{c}{$RNDs from SEI-American Exercise$}&\multicolumn{1}{c}{$$}\\
%RNDs from SEI 
SEI &T.FTSE100 v Median &0.955 & 1.01 & 0.05 &Accept\\
SEI &Realised Vol. v Ann.I.V.$\%$ & 0.318 & 3.58 & -2.55 &Reject\\
% &\multicolumn{3}{c}{$Market Comparison$}&\multicolumn{1}{c}{$$}\\
%Market Comparison
Market Outcome &Realised Vol. v ATMIV(LIFFE)$\%$ & 0.426 & 1.46 & -4.50 &Reject\\
\end{tabular}
\end{ruledtabular}
\footnotetext[1]{$F_{crit} =1.54.\hspace{3 pt} t_{crit}(2 \hspace{3 pt} tail) =2.00.\hspace{3 pt} H0:=no \hspace{3 pt} difference \hspace{3 pt} in \hspace{3 pt} means \hspace{3 pt} at \hspace{3 pt} 95\% \hspace{3 pt} confidence.  $} %\footnotetext[2]{$t_{crit}(2 \hspace{6 pt} tail) =2.00.$}
\normalsize
\end{table}
Further tests were carried out to assess the practical effects of these differences on the predictive properties of RNDs from each type of put option. Results of these tests are presented in Table II. The median and annualised percentage implied volatility from each RND are compared with the actual traded FTSE 100 closing price (T.FTSE 100) and realised volatility on the expiration date of the option. This is a test of the one month (17/18 trading days) predictive characteristics of the median and implied volatility from the estimated RNDs. The t-test results in Table II indicate that the medians of RNDs from both American and European exercise options provide an unbiased estimate of FTSE 100 closing prices on the expiration date of the options, one month  later. The annualised implied volatilities of the RNDs from both types of options, on the other hand, are biased estimates of the actual realised volatilities at expiration of the option. However, the volatility estimates from the RNDs compare favourably with LIFFE tabulated at-the-money implied volatility, which gives an even more biased estimate of realised volatility.  These results suggest unbiased estimates of future asset prices can be obtained from RNDs from both European and American put options on those assets.
\begin{table}
\scriptsize
\caption{\label{tab:table3}Skewness and Kurtosis: Estimated RNDs v Normal.}
\begin{ruledtabular}
\begin{tabular}{cccccccc}
 $Option$ &$Skewness$
 &$Normal$  &$t-stat(paired)$\footnotemark[1] & &$Kurtosis$
 &$Normal$  &$t-stat(paired)$\footnotemark[1]\\
\hline
ESX & -0.43 & 0 &-10.1 & & 2.82 & 3 & -4.4\\
SEI & -0.65 & 0 &-14.7 & & 3.19 & 3 & 3.4\\
\end{tabular}
\end{ruledtabular}
\footnotetext[1]{Reject $H0:=no \hspace{3 pt} difference \hspace{3 pt} in \hspace{3 pt} means  \hspace{3 pt} at \hspace{3 pt} 95\% \hspace{3 pt} confidence$ if $abs(t-test)$ $ >$  $t_{crit}(2 \hspace{3 pt} tail) =2.00.$}
\normalsize
\end{table}
Table III shows the average departure of the ESX and SEI RNDs from normality. The t-stats from paired t-tests of  the series of values of skewness and kurtosis estimated from the RNDs, and those for the normal distribution, lead to rejection of the null hypothesis of no difference at the 95$\%$ level.  It is important to appreciate that RNDs extracted from option prices are distinct from the historical distribution of the underlying asset returns (prices), since one is risk adjusted and the other is not. However, their shapes should be the same.  Thus, if asset returns were normally distributed we would not expect to obtain negatively skewed, leptokurtic, or platykurtic RNDs.  The results above are therefore consistent with other empirical findings suggesting that asset returns are not normally distributed.   

\section{Discussion  and Conclusions}
In this empirical study, we applied our method of estimating RNDs to a large set of FTSE 100 European style put option daily data, and then as an Ansatz to a corresponding set of American style options on the same underlying asset. Our results in Table I suggest that the RNDs obtained from each style of option are significantly different, reflecting the distorting influence of the early exercise possibilities for the American put options. We confirmed that estimates of volatility from the RNDs from both types of option were biased estimates of the actually realised volatilities at expiration, suggesting that prices for the options tend to over estimate volatility. However, caution is necessary in interpreting the latter results, as it is difficult to reliably estimate realised volatility from at most 18 daily observations of returns. The values of skewness and kurtosis obtained also suggest that the underlying pricing process departs from geometric Brownian motion. The results presented in Table II are surprising.  They suggest that in practice, RNDs from both European and American put options can be used interchangeably to obtain estimates of future asset prices. This holds although the theory underlying RNDs outlined in section II applies only to European style exercise, and despite the existence of significant model differences revealed by the (albeit more powerful) paired t-test results presented in Table I. In addition, the standard deviation of RNDs for the American put options is smaller at 107 index points compared with 113 on average for the European options.  Reassuringly, FIG. 1 shows for ESX how in all cases the actual FTSE 100 closing prices lie within U and L, the $\pm$2 standard deviation confidence intervals constructed from the estimated standard deviations of each separate RND; the same applies to SEI. Overall, our results suggest that neural nets provide a promising method for use in extracting RNDs from option prices; this merits further investigation. To evaluate fully the potential and limitations of  the approach we describe here, an empirical comparison of the statistical characteristics of RNDs from the double lognormal, from a smoothed implied volatility smile, and from our approach using neural nets is required. In particular, further consideration needs to be given to the observation of Dupont [12] on deep in-the-money American style options.

\begin{figure*}
\includegraphics{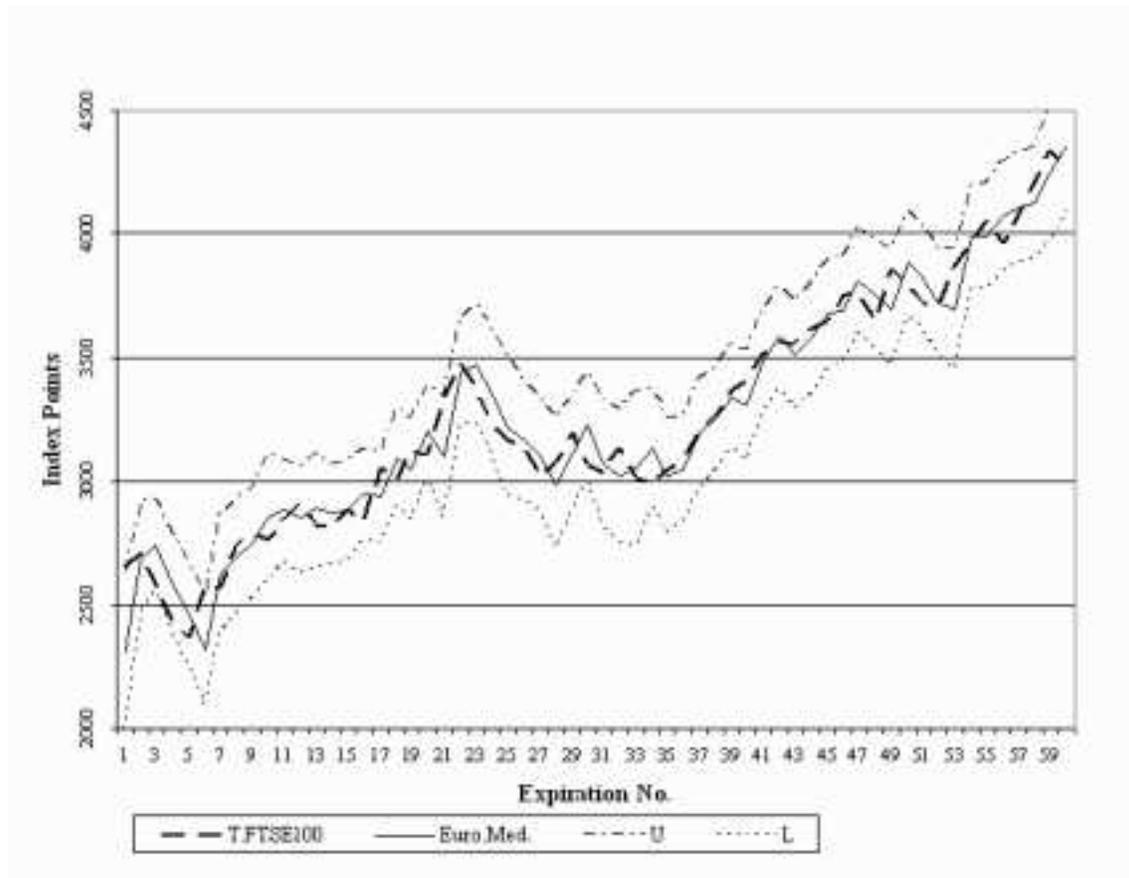}% Here is how to import EPS art
\caption{\label{fig:wide}ESX Put Option RNDs: One Month Estimate of Level of FTSE 100 showing it lies within the 95$\%$ confidence band of the model.}
\end{figure*}

%\begin{figure*}
%\includegraphics{figure2}% Here is how to import EPS art
%\caption{\label{fig:wide}SEI Put Option RNDs: One Month Estimate of Level of FTSE 100.}
%\end{figure*}

%Fig.1 and 2 give one month estimates of the FTSE 100 closing price from RNDs for European and American options respectively. The estimate values are the medians of the RNDs. The heavy black dashed lines are the FTSE 100 closing price one month (17/18 trading days) later.  The dotted lines are $\pm$2 standard deviation ($\pm$95.46$\%$) confidence intervals. The true values lie within the confidence band in all cases. 

%\newpage %Just because of unusual number of tables stacked at end
%\bibliography{RND}% Produces the bibliography via BibTeX.

\end{document}